\documentclass[%
 reprint,
 amsmath,amssymb,
 aps,
]{revtex4-1}

\usepackage{graphicx}% Include figure files
\usepackage{dcolumn}% Align table columns on decimal point
\usepackage{bm}% bold math
\usepackage{dsfont}
\usepackage{appendix}
\usepackage{color}
\usepackage{ulem}

\begin{document}

\preprint{APS/123-QED}

\title{A solution for the cosmological constant problem}
% Force line breaks with

\author{Renata Jora
	$^{\it \bf a}$~\footnote[1]{Email:
		rjora@theory.nipne.ro}}
\email[ ]{rjora@theory.nipne.ro}

\affiliation{$^{\bf \it a}$ National Institute of Physics and Nuclear Engineering PO Box MG-6, Bucharest-Magurele, Romania}

\begin{abstract}

We consider that the cosmological constant is associated with the vacuum energy density of a particle physics model.  In the path integral formalism of euclidean quantum gravity and in the background of the Robertson Walker metric we calculate only the contribution of the matter density to the fluctuating metric.  This produces an effective term in the action that acts as a correction to the vacuum energy density of the matter. The method leads to the observed cosmological constant without the need of any fine tuning in the gravity or particle physics sectors.

\end{abstract}
\maketitle

\section{Introduction}

Over the years cosmological observation \cite{Peebles}-\cite{Turner1} have indicated that there is in the Universe a homogeneous energy density that leads to the accelerating expansion of the Universe.  This energy density is called the cosmological constant. 
Latest astrophysical data  from type I supernovae \cite{Hansen}-\cite{Garnavich2}, from cosmic microwave background \cite{Spergel1}-\cite{Melchiorri}, matter density  \cite{Schramm}-\cite{Lewis} and gravitational lensing \cite{Gott}-\cite{Turner3} to enumerate only some of the results brought even more support to the claim that a small cosmological constant could be the only possible explanation compatible to our experimental knowledge.

The main theoretical problem associated with the cosmological constant stems from its association with the vacuum energy of the quantum field theory of the matter in the gravitational field. No matter the scale at which the QFT is considered or the values of the condensates which are taken into account the corresponding energy density is too large by enormous orders of magnitude as compared to the observed one. For example \cite{Carroll} if the scale of interest is the electroweak scale the vacuum density should be of the order:
\begin{eqnarray}
\rho_v^{EW}\approx (200\,\, GeV)^4,
\label{vac554663}
\end{eqnarray}
as compared to the observed one:
\begin{eqnarray}
\rho_{\Lambda}\approx (10^{-12}\,\,GeV)^4.
\label{observed54}
\end{eqnarray}

A possible explanation of the small cosmological constant can be found not without fine-tuning in supersymmetry \cite{Niles}-\cite{Cremmer}, string theory \cite{Dine}-\cite{Banks} or even in the context of the anthropic principle \cite{Weinberg1}-\cite{Garriga}. Other possible scenarios that may solve this issue based on modified theories of gravity or considering exotic quantum effects are listed in \cite{Press}-\cite{Weinberg2}.

\section{A solution for the cosmological constant}

The Einstein action in general relativity is given by:
\begin{eqnarray}
S=\int d^4 x \sqrt{-g}\Bigg[\frac{1}{2{\it k}}(R-2\Lambda)+{\cal L}_m\Bigg],
\label{action6577}
\end{eqnarray}
where $R$ is the curvature, $\Lambda$ is the cosmological constant and ${\cal L}_M$ is the Lagrangian of matter. The extremum of the action is realized by considering $\frac{\delta S}{\delta  g^{\mu\nu}}$ and this leads to the  Einstein equations:
\begin{eqnarray}
R_{\mu\nu}-\frac{1}{2}g_{\mu\nu}R+\Lambda g_{\mu\nu}=\frac{8\pi G}{c^4}T_{\mu\nu}.
\label{einsteq8677}
\end{eqnarray}
Here $R_{\mu\nu}$ is the Ricci tensor, $T_{\mu\nu}$ is the matter energy momentum tensor and $G$ is the gravitational constant. 

The cosmological evolution of the Universe  which homogenous and isotropic  is described by the Robertson Walker  (RW) metric \cite{PDG}:
\begin{eqnarray}
ds^2=dt^2-R^2(t)\Bigg[\frac{d r^2}{1-kr^2}+r^2(d\theta^2+\sin^2(\theta)d\Phi^2\Bigg].
\label{Rwmetric756}
\end{eqnarray}
In the above equation $k$ may take three values $+1$, $-1$, $0$ corresponding to closed, open or flat geometries.

The Hubble parameter has the expression:
\begin{eqnarray}
H(t)=\frac{\dot{R}(t)}{R(t)}.
\label{hubble65664}
\end{eqnarray}
Furthermore one can define present day density parameter for the cosmological constant as $\Omega_{\Lambda}=\frac{\Lambda}{3H^2}$. The astrophysical data \cite{Riess1}-\cite{Betoul} indicates that $\Omega_{\Lambda}=0.685\pm0.007$ at $k=0$ \cite{PDG}. This would correspond to $\Lambda\approx 4. 33*10^{-66} {\rm eV}^2$.

 We first integrate in the path integral formalism over the quantum matter degrees of freedom to obtain:
\begin{eqnarray}
Z_m=\exp[-\int d^4 x\rho\sqrt{-g}]=
\int d\Phi_i \exp[-d^4x {\cal L}_m],
\label{act45536}
\end{eqnarray}
where $\Phi_i$ are all matter degrees of freedom, scalars, fermions  or gauge bosons.

Consider euclidean quantum gravity in the background of the RW metric $g_{\mu\nu}\rightarrow t_{\mu\nu}+h_{\mu\nu}$ where $h_{\mu\nu}$ is the RW metric and $t_{\mu\nu}$ is the fluctuating metric to be integrated over in the partition function. We disregard our lack of clear knowledge of a quantum theory of gravity. We first expand $\sqrt{-g}$ in quadratic order which yields:
\begin{eqnarray}
&&\sqrt{-\det[g_{\mu\nu}]}=\sqrt{-\det [h_{\mu\nu}]}\Bigg[1+\frac{1}{2}h^{\rho\sigma}t_{\rho\sigma}-
\nonumber\\
&&\frac{1}{4}h^{\rho\sigma}t_{\sigma\alpha} h^{\alpha\beta}t_{\beta\rho}+\frac{1}{8}h^{\rho\sigma}t_{\rho\sigma}h^{\alpha\beta}t_{\alpha\beta}\Bigg].
\label{finres43552}
\end{eqnarray}
Then we consider the gravitational partition function as follows:
\begin{eqnarray}
&&Z_g=\int dt_{\mu\nu} \mu(t_{\mu\nu}) \exp\Bigg[-\int d^4 x \sqrt{-g}\frac{1}{2{\it k}}R(g)\Bigg]\times
\nonumber\\
&&\exp\Bigg[-\int d^4 x\rho\sqrt{-\det [h_{\mu\nu}]}[1+\frac{1}{2}h^{\rho\sigma}t_{\rho\sigma}-
\nonumber\\
&&\frac{1}{4}h^{\rho\sigma}t_{\sigma\alpha} h^{\alpha\beta}t_{\beta\rho}+\frac{1}{8}h^{\rho\sigma}t_{\rho\sigma}h^{\alpha\beta}t_{\alpha\beta} ]\Bigg]=
\nonumber\\
&&\int dt_{\mu\nu} \mu(t_{\mu\nu})\exp\Bigg[-\int d^4 x \sqrt{-h}\frac{1}{2{\it k}}R(h)\Bigg]\times
\nonumber\\
&&[1+....{\rm corrections \,\,to\,\, R(h)}]\times
\nonumber\\
&&\exp\Bigg[-\int d^4 x\rho\sqrt{-h}[1+\frac{1}{2}h^{\rho\sigma}t_{\rho\sigma}-
\nonumber\\
&&\frac{1}{4}h^{\rho\sigma}t_{\sigma\alpha} h^{\alpha\beta}t_{\beta\rho}+\frac{1}{8}h^{\rho\sigma}t_{\rho\sigma}h^{\alpha\beta}t_{\alpha\beta} ]\Bigg].
\label{gravit66588}
\end{eqnarray}
Here $\mu(t_{\mu\nu})$ is a measure that appears in the path integral formalism (this can be introduced in the exponent and then  consider only the zeroth order contribution). Then,
\begin{eqnarray}
&&Z_g=\int dt_{\mu\nu} \exp\Bigg[-\int d^4 x \sqrt{-h}\frac{1}{2{\it k}}R(h)\Bigg]\times
\nonumber\\
&&[1+....{\rm full\,\,corrections \,\,to\,\, R(h)}]\times
\nonumber\\
&&\exp \Bigg[-\int d^4 x\rho\sqrt{-h}[1+\frac{1}{2}h^{\rho\sigma}t_{\rho\sigma}-
\nonumber\\
&&\frac{1}{4}h^{\rho\sigma}t_{\sigma\alpha} h^{\alpha\beta}t_{\beta\rho}+\frac{1}{8}h^{\rho\sigma}t_{\rho\sigma}h^{\alpha\beta}t_{\alpha\beta} ]\Bigg].
\label{gravit6658895}
\end{eqnarray}

We integrate over all degrees of freedom $t_{\mu\nu}$ making the assumption that eventual constraints that diminish the number of degrees of freedoms that appear in the form of gauge fixing terms are disregarded. Moreover we do not take into account any symmetry of the metric and regarded as such, a two indices tensor with $16$ components.
\begin{eqnarray}
&&Z_g\approx \exp\Bigg[-\int d^4 x \sqrt{-h}[\frac{1}{2{\it k}}R(h)+\int d^4 x\rho]\Bigg]\times
\nonumber\\
&&\det\Bigg[[-\frac{1}{4}h^{\rho\sigma}t_{\sigma\alpha} h^{\alpha\beta}t_{\beta\rho}+\frac{1}{8}h^{\rho\sigma}t_{\rho\sigma}h^{\alpha\beta}t_{\alpha\beta} ](-\rho)\Bigg]^{-1/2}.
\label{rezhsgdf}
\end{eqnarray}
Note that we did not extract the factor $\sqrt{-h}$ as  this is regarded as a property of the curved lattice one may use for this space. We leave aside the first big factor in the square bracket of the second line of Eq. (\ref{rezhsgdf}) and we consider only the contribution from $\rho$. This leads to:
\begin{eqnarray}
&&[\det[\rho]]^{-1/2}=\exp[{\rm Tr}-\frac{1}{2}\ln[\rho]]=
\nonumber\\
&&\exp\Bigg[-d^4x \sqrt{-h}\frac{16}{2}\int \frac{d^4 p}{(2\pi)^4}\ln[\rho]\Bigg],
\label{res553442}
\end{eqnarray}
where the trace is considered over the curved space and all $16$ degrees of freedom of the tensor $t_{\mu\nu}$.

We consider  the contribution of matter from both the exponent in Eq. (\ref{rezhsgdf}) and from Eq. (\ref{res553442}) and Eq. (\ref{res553442}) to determine:
\begin{eqnarray}
&&Z_g=\exp\Bigg[-\int d^4 x \sqrt{-h}[\frac{1}{2{\it k}}R(h)+
\nonumber\\
&&\rho+ \int \frac{ d^4 p}{(2\pi)^4}8\ln(\rho)]\Bigg].
\label{finalres75664}
\end{eqnarray}

The Einstein equations derived form Eq. (\ref{finalres75664}) are then:
\begin{eqnarray}
R_{\mu\nu}-\frac{1}{2}h_{\mu\nu}R+\Lambda h_{\mu\nu}=\frac{8\pi G}{c^4}T_{\mu\nu},
\label{res546354}
\end{eqnarray}
where in our approach,
\begin{eqnarray}
\Lambda=\frac{8\pi G}{c^4}[\rho+8\int \frac{d^4 p}{(2\pi)^4}\ln(\rho)].
\label{final657888}
\end{eqnarray}

We use $M_P^2=\frac{\hbar c}{G}$ where $M_P\approx 1.22\times 10^{19}$ GeV and natural units. 

We introduce $M$ as the ultraviolet scale of the theory in the quantum field theoretical approach. One may  write $\rho\approx y M$ where $y$ should be a dimensionless factor with sensible value that is not fine-tuned. Then Eq. (\ref{final657888}) becomes:
\begin{eqnarray}
\Lambda=8\pi \frac{M^4}{M_P^2}[y+\frac{1}{2\pi^2}\ln(y)],
\label{alm86776}
\end{eqnarray}
where $\int \frac{d^4 p}{(2\pi)^4}=\frac{M^4}{16\pi^2}$ and the logarithm is written  as $\ln(\frac{\rho}{M^4})$.

We consider first $M=M_P$ the ultraviolet scale at the Planck scale and second $M=200$ GeV at the electroweak scale to solve Eq. (\ref{alm86776}) with the value of $\Lambda$ taken from astrophysical data $\Lambda \approx 4. 33*10^{-66}{\rm eV}^2$. We find that for both cases $y\approx 0.111$ and $y^{1/4} \approx 0.578$ which shows that this value is not tuned. Therefore no matter the ultraviolet scale or the actual value of the QFT vacuum of the theory such an approach may lead to the correct cosmological constant.

\section{Conclusions}

In the background gauge field method one couples the gauge theory to a classical background gauge field and integrates in the partition function over quantum fluctuations.  A quantum theory that incorporates gravity should be no exception.  For cosmological theories one just needs to couple the quantum theory with the background classical RW metric and integrate over quantum fluctuations of the metric. The problem is that we do not have as yet an established theory of quantum gravity. 

In this work we show how one might circumvent this problem in computing corrections to the cosmological constant.  This starts from the observation that in order to compute gravity corrections to the vacuum density of a matter quantum field theory  one does not really need to know how to integrate the metric but only to lay down an adequate path integral formalism of gravity with many unknowns (like the actual measure, the constraints, the gauge fixing or the boundary conditions).  This is because the actual gravitational contributions can be separated in the formalism from the matter contributions and put in separate terms that can be ignored at least in the first order.  It turns out that this approach can lead to a correct description of the cosmological constant without any need of fine tuning. This is put in evidence by Eq. (\ref{alm86776}) in the present paper.

The conclusions is that a quantum gravitational approach to the contribution of matter density can produce the observed cosmological constant without any need to alter to particle physics theories that are known and established.  The approach here is straightforward and well defined and eludes  the necessity of an accepted theory of quantum gravity. Therefore it is justified from all points of view.

\end{document}